\documentclass[twocolumn,prl,psfig,superscriptaddress]{revtex4}
\usepackage{graphicx}
\usepackage{epsfig}
\begin{document}
\title{Magnetic anisotropy of BaCu$_2$Si$_2$O$_7$: theory and antiferromagnetic resonance
 }

\author{R. Hayn}
\affiliation{Laboratoire Mat{\'e}riaux et Micro{\'e}lectronique de
Provence associ{\'e} au CNRS, Facult{\'e} de Saint-J{\'e}r{\^o}me,
13397 Marseille Cedex 20, France}

\author{V.A. Pashchenko}
\affiliation{ Institute for Low Temperature Physics and
Engineering, National Academy of Sciences of Ukraine, 310164
Kharkov, Ukraine}

\author{A. Stepanov}
\affiliation{Laboratoire Mat{\'e}riaux et Micro{\'e}lectronique de
Provence associ{\'e} au CNRS, Facult{\'e} de Saint-J{\'e}r{\^o}me,
13397 Marseille Cedex 20, France}

\author{T. Masuda}
\affiliation{Department of Advanced Materials Science, The University of
Tokyo, 6th Engineering Bldg., 7-3-1 Hongo, Bunkyo-ku, Tokyo, 113-8656, Japan}
\author{K. Uchinokura}
\affiliation{Department of Advanced Materials Science, The University of
Tokyo, 6th Engineering Bldg., 7-3-1 Hongo, Bunkyo-ku, Tokyo, 113-8656, Japan}
\date{\today}
%\maketitle

\begin{abstract}
Antiferromagnetic resonance (AFMR) of BaCu$_2$Si$_2$O$_7$ and
a microscopic theory for the magnetic anisotropy of spin 1/2 chain
compounds with folded CuO$_3$ geometry being in good agreement with 
the available data are presented. 
The AFMR studies at 4.2 K show the existence of two gaps (40 and 76 GHz)
at zero magnetic field and of two spin re-orientation transitions for 
$H \parallel c$. The microscopic origin of the two gaps is shown to be 
Hund's rule coupling which leads to a ``residual anisotropy'' beyond the 
compensation of the Dzyaloshinskii-Moriya term by the symmetric 
anisotropy which would be valid without Hund's coupling.
%Both
%compounds contain folded CuO$_3$ chains.
%The theory treats the transfer terms,
%the spin-orbit interaction and Hund's exchange like a perturbation in a
%Cu-O-Cu cluster. The deviation from a straight bond leads to a
%Dzyaloshinskii-Moriya (DM) interaction and a spin canting angle. This DM
%term is compensated by a symmetric anisotropy term as it was also found by
%Aharony for other cuprate compounds. Beyond this compensation, there is a
%``residual anisotropy'' due to
%Hund's exchange which leads to an easy axis anisotropy.
%That ``residual anisotropy'' is very small of the order of 5 $\mu$eV only.
%The DM terms, the compensating anisotropy terms and the ``residual
%anisotropy'' estimated within the present theory are not only in qualitative
%but also in quantitative agreement with available magnetization and neutron
%scattering data for both compounds.
\end{abstract}
\maketitle

There exist several cuprate compounds with CuO$_3$ corner-sharing
chains that are known as nearly ideal model compounds for
one-dimensional spin 1/2 systems. Those with folded CuO$_3$ zigzag
chains, like BaCu$_2$Ge$_2$O$_7$ (N{\'e}el temperature of
$T_N=8.8$ K) \cite{tsukada00} or BaCu$_2$Si$_2$O$_7$ ($T_N=9.2$ K)
\cite{tsukada99,kenzelmann01,tsukada01,zheludev01} open the
fascinating possibility to study noncollinear magnetism or the
influence of a Dzyaloshinskii-Moriya (DM) term in a quantum spin
chain. Especially BaCu$_2$Si$_2$O$_7$ attracted much interest
recently in connection with the finding of two consecutive spin
re-orientation transitions for a magnetic field applied along the
easy $c$-axis, which were discussed controversially, however
\cite{tsukada01,zheludev01}. Furthermore, one \cite{tsukada99} or
two gaps \cite{kenzelmann01} in the spin-wave spectrum measured by
neutron scattering were reported. And also the microscopic origin
of the magnetic interaction energies was not clarified up to now.
The present work addresses all these points by a combined
experimental and theoretical study, using antiferromagnetic
resonance (AFMR) measurements and perturbation theory with respect to
spin-orbit (SO) coupling in a Cu-O-Cu cluster with folded
geometry.

Our experimental study shows clearly the existence of two
different gaps at zero magnetic field in BaCu$_2$Si$_2$O$_7$ and a
magnetic structure evolution with increasing field $H\parallel c$
with two re-orientation transitions corresponding to a spin
rotation towards the middle $b$-axis followed by a rotation
towards the hard $a$-axis which confirms the neutron scattering
structure of Ref. \cite{zheludev01}.

Theoretically we show that the deviation from a straight bonds in
a Cu-O-Cu cluster leads to a DM interaction and a spin canting
of neighboring spins at $T<T_N$. The canting angle is rather large
and the DM interaction alone would lead to much larger gaps in
spin wave spectrum and AFMR than actually observed. It is
the symmetric anisotropy part which compensates the DM
one, as proposed in Refs.\ \cite{kaplan83,shekhtman92} and applied
here to the given situation. 
A clear experimental confirmation of this effect was found in 
Ba$_2$CuGe$_2$O$_7$ \cite{zheludev98}.
We argue that Hund's coupling leads
to a ``residual anisotropy'', beyond this compensation,  which
agrees well with the experimental data and explains the observed
easy axis behavior and the two gaps mentioned above. Naturally,
our theory is also valid for BaCu$_2$Ge$_2$O$_7$ which differs
from BaCu$_2$Si$_2$O$_7$ by the interchain couplings. 
\begin{figure}[b]
\includegraphics[width=9cm,angle=0]{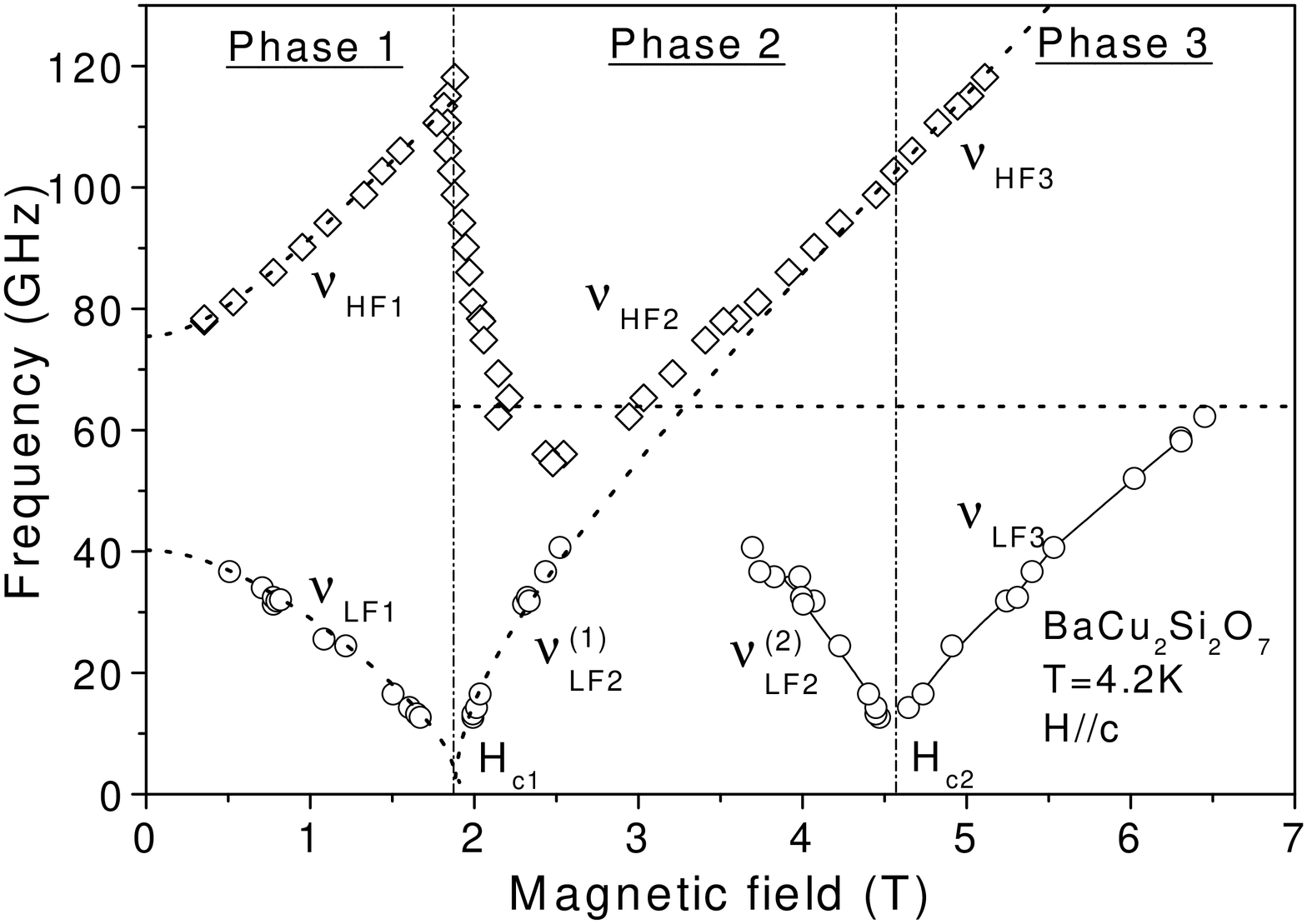}
\caption{ The frequency vs field diagram ($H \parallel c$) of  AFMR modes in
BaCu$_2$Si$_2$O$_7$ at $T=4.2$ K. Dashed lines: AFMR modes of a
biaxial antiferromagnet without the DM interaction in a collinear
(phase 1) and a canted (phase 2) phase according to Ref.\  
\cite{nagamiya}.}
\end{figure}
The more
simple AFM structure of the Ge-compound allows for a weak
ferromagnetic moment which gives direct information on the spin
canting angle \cite{tsukada00}.

The crystal structure of BaCu$_2$Si$_2$O$_7$  and
BaCu$_2$Ge$_2$O$_7$ belonging to the orthorhombic space group
$Pnma$ is  made of almost isolated CuO$_3$ corner-sharing
chains running along the $c$-axis \cite{oliviera}. The Cu-O-Cu
bond angle is found to be 124$^{\circ}$ in the case of
BaCu$_2$Si$_2$O$_7$ which is smaller than that of BaCu$_2$Ge$_2$O$_7$
(135$^{\circ}$). For both compounds a broad maximum was observed
in the temperature dependence of the magnetic susceptibility
indicating 1D magnetic behavior \cite{tsukada99} and allowing to 
extract the interchain exchange constants $J=47$ meV ($J=24$ meV)
for the Ge (Si) compound.
The single crystals of BaCu$_2$Si$_2$O$_7$ used in an
AFMR experiment were grown by a
floating-zone method. Since the $a$- and $c$- axis lengths are
almost the same, particular attention was paid to the crystal
orientation. The AFMR study of BaCu$_2$Si$_2$O$_7$ was done using
a simple millimeter-range video spectrometer \cite {zvyagin85}.
The microwave sources were either Gunn's diodes  or back-ward wave
tubes.  The field dependence of AFMR modes in BaCu$_2$Si$_2$O$_7$
at $T=4.2$ K is shown in Fig.\ 1. Two gaps are clearly seen at $H=0$,
$\nu_{LF}=40$ GHz (low-frequency mode) and $\nu_{HF}=76$ GHz
(high-frequency mode). The gaps values are about 10\% smaller as
compared to the earlier reported values 0.21 meV  and  0.36 meV
measured at lower temperature $T=2$ K \cite{kenzelmann01}. The
observed frequency softening  of the low-frequency mode  at
$H_{c1}$ and $H_{c2}$ confirms the existence, for this particular
magnetic field orientation, of two consecutive spin re-orientation
transitions. As it  was first proposed in Ref. \cite{zvyagin85}
this behavior is characteristic of a AFM with the DM interaction
in a  magnetic field applied along the easy axis and the DM vector
orientation along the middle axis ($c$- and $b$-axis respectively
in  the case of BaCu$_2$Si$_2$O$_7$; see Fig.\ 4 of Ref.\ 
\cite{zvyagin85} for comparison). Furthermore, our observation of quadratic
field dependence of the high-frequency AFMR mode for $H\parallel
a$ (not presented in Fig.\ 1) allows the conclusion that the $a$-axis is the
hard axis of the magnetic anisotropy
tensor of BaCu$_2$Si$_2$O$_7$ \cite{remark5}.

To understand the microscopic origin of the observed behavior in
Fig.\ 1 and also other experimental facts already established for
the two compounds in question we investigate the special Cu-O-Cu
bond sketched in Fig.\ 2 with an arbitrary bond angle $\pi -
\phi$. The isotropic exchange, the DM term and the symmetric
anisotropy are calculated by perturbation theory with respect to
SO coupling and kinetic energy. The Hamiltonian contains
all
the 3$d$ orbitals at the 2 Cu-sites (denoted by $m$ or $m' = \{
0 (d_{x^2-y^2}), z (d_{xy}), x (d_{xz}), y (d_{xz}), 1 (d_{3z^2-r^2}) \}$,
where the usual notation is given in parenthesis) and the three 2$p$ orbitals
($n$ or $n' = \{ p_x, p_y, p_z \}$) at the oxygen site
in between. The local,
unperturbed part contains the Coulomb terms
%\begin{widetext}
\begin{eqnarray}
\hat{H}_0 &=&
E^p n^p + \frac{U_p}{2} \left(n^p n^p - n^p \right)
\nonumber \\
&+& \sum_{\alpha m} E_m n^d_{\alpha m}
+ \frac{U_d}{2} \sum_{\alpha}
\left( n^d_{\alpha} n^d_{\alpha} - n^d_{\alpha} \right) \; , 
%\sum_{n \sigma} E^p p_{n \sigma}^+ p_{n \sigma}
%+ \frac{U_p}{2} \sum_{ n n' \sigma \sigma'}
%p_{n \sigma}^+ p_{n' \sigma'}^+
%p_{n' \sigma'} p_{n \sigma}
%\nonumber \\
%&& + \sum_{\alpha m \sigma} E_m
%d_{\alpha m \sigma}^+ d_{\alpha m \sigma}
%+ \frac{U_d}{2} \sum_{\alpha m m' \sigma \sigma'}
%d_{\alpha m \sigma}^+ d_{\alpha m' \sigma'}^+
%d_{\alpha m' \sigma'} d_{\alpha m \sigma}
\label{1}
\end{eqnarray}
%\end{widetext}
with $n^d_{\alpha}=\sum_m n^d_{\alpha m} =
\sum_{m \sigma} d^{\dagger}_{\alpha m \sigma} d_{\alpha m \sigma}$
and $n^p=\sum_{n \sigma} d^{\dagger}_{n \sigma} p_{n \sigma}$ and
where $\alpha=A$ or $B$ (sort of Cu),
%$m$ and $m'$ are $d$-orbitals, $n$ and
%$n'$ denote $p$-orbitals
and $\sigma$ ($\sigma'$) are spin indices. The charge
transfer energy $\Delta_p=E^p-E_0$ has been set to 4 eV \cite{hybertsen}
and $E_{m}-E_0=\varepsilon_d=2$ eV shall be the same for all $m 
= \{x, y, z, 1 \}$ \cite{remark2}. We use in the following the standard
parameters of cuprates, i.e.\ $U_p=4$ eV and $U_d=10$ eV.
%Since there are at present no band structure calculations
%available for the compounds of interest, we take a value of $\varepsilon_d$
%that is known for the
%related cuprate Ba$_3$Cu$_2$O$_4$Cl$_2$ \cite{yushankhai}.
All the remaining
terms
of the Hamiltonian are treated as a perturbation. That concerns especially the
kinetic energy
\begin{equation}
\label{2}
\hat{H}_t=\sum_{\alpha m n \sigma} t_{\alpha m, n}
(d_{\alpha m \sigma}^{\dagger} p_{n \sigma} + h.c. ) \; , 
\end{equation}
and the spin-orbit interaction
\begin{equation}
\label{3}
\hat{H}_{SO}= \frac{\lambda}{2}\sum_{b \alpha} \sum_{m m' \sigma \sigma'}
d_{\alpha m' \sigma'}^{\dagger} L_{m m'}^b \sigma_{\sigma \sigma'}^b
d_{\alpha m \sigma} \; , 
\end{equation}
($\lambda=0.1$ eV)
with the only nonzero matrix elements $L_{z0}^z=2i$, $L_{x0}^x=-i$, and
$L_{y0}^y=-i$ ($L_{m m'}^b=(L_{m' m}^{b})^{*}$), where $b$ is a Cartesian
coordinate and $\sigma_{\sigma \sigma'}^b$ the corresponding Pauli spin
matrix. The Hund's exchange interaction 
%(that will be shown to be responsible
%for the ``residual anisotropy'') 
is given by
\begin{eqnarray}
\label{4}
\hat{H}_H &=& -
J_{Hd} \sum_{\sigma \sigma'} \sum_{\alpha m m'}^{m\neq m'}
d_{\alpha m \sigma}^{\dagger} d_{\alpha m \sigma'}
d_{\alpha m' \sigma'}^{\dagger} d_{\alpha m' \sigma}
\nonumber \\
&& - J_{Hp} \sum_{\sigma \sigma'} \sum_{n n'}^{n\neq n'}
p_{n \sigma}^{\dagger} p_{n \sigma'}
p_{n' \sigma'}^{\dagger} p_{n' \sigma} \; . 
\end{eqnarray}
\begin{figure}[t]
\includegraphics[width=7.5cm,angle=0]{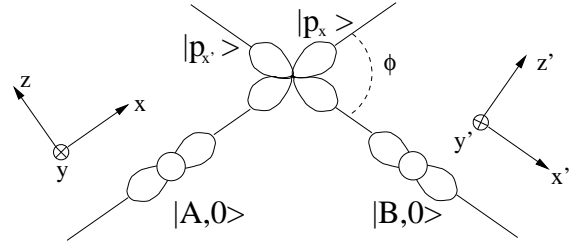}
\caption{
The orbitals of the Cu$_2$O cluster and the coordinate systems used in the
calculation. The two ground state orbitals $3d_{x^2-y^2}$ and
$3d_{x'^2-y'^2}$ are fixed by the positions of the two CuO$_4$
plaquettes perpendicular to the plane of the Figure. }
\end{figure}
%The formulation chosen here is slightly different from Ref.\ \cite{aharony98}
%and allows to incorporate Hund's exchange interaction more easily.
%But we have checked that both formulations give exactly the same answers.
In a first step, the intermediate oxygen orbitals are excluded and an 
effective spin Hamiltonian
%
%We now formulate the general theory before we specify it to the present
%situation. The formulation given by Aharony \cite{aharony}
%(see also \cite{yushankhai}) starts by deriving an effective $d$-$d$ transfer
%including the SO interaction (\ref{3}). That effective interaction is then
%treated in second order to obtain DM and symmetric anisotropy terms, and to
%see its compensation immediately. Here, we
%would like to give a more explicit formulation into which Hund's exchange
%interaction can be incorporated more easily. It should be
%noted, however, that we have checked that both formulations give exactly the
%same
%answers. In a first step, we exclude the intermediate oxygen orbitals (the
%$\hat{H}_t$ term) and derive an effective spin Hamiltonian between the two
%Cu sites
\begin{eqnarray}
\label{5}
\hat{H}^{ex}&=&\frac{1}{2}
\sum_{\{m\}\sigma\sigma'}
d_{A m_1 \sigma}^{\dagger} d_{A m_3 \sigma'}
d_{B m_2 \sigma'}^{\dagger} d_{B m_4 \sigma}
\nonumber \\
&& \cdot
J({m_1}_A,{m_2}_B;{m_3}_A,{m_4}_B)
\end{eqnarray}
is derived. The necessary exchange terms are calculated in 4th order
perturbation theory with respect to the kinetic energy $\hat{H}_t$.
%That interaction contains for example the isotropic exchange
%$J=J(0_A,0_B;0_A,0_B)$.
The magnetic anisotropy terms arise due to the
SO interaction. Up to second order, the result can be written as
%In first order, we obtain the DM interaction
%(if it is allowed by symmetry).
%The second order corresponds to
%a symmetric spin anisotropy. Both terms together can be written as
%\begin{widetext}
\begin{equation}
\hat{H}_{DM}+\hat{H}_A = \vec{D} (\vec{S}_A \times \vec{S}_B )
+ \vec{S}_A \vec{\vec{\Omega}} \vec{S}_B \; , 
\end{equation}
with
$\Omega^{ab}=\Gamma^{ab}+\Gamma^{ba}-\delta^{ab} (\sum_c \Gamma^{cc})$.
The first order term corresponds to a DM interaction:
\begin{equation}
D^b = \sum_m \frac{i \lambda L_{0m}^b
( J(0_A,m_B;0_A,0_B)-J(m_A,0_B;0_A,0_B) ) }
{E_m-E_0}
\end{equation}
and the second order describes a symmetric spin anisotropy
\begin{eqnarray}
\Gamma^{ab} &=& \left( \frac{\lambda}{2} \right)^2
\sum_{m m'} \left\{
\frac{L_{0m}^a L_{m'0}^b J(m_A,0_B;m'_A,0_B)}
{(E_m-E_0)(E_{m'}-E_0) } \right.
\nonumber \\
&&+ \left. \frac{L_{0m}^a L_{0m'}^b J(m_A,m'_B;0_A,0_B)}
{(E_m-E_0)(E_{m'}-E_0)}
+ A \leftrightarrow B \right\} \; .
\nonumber
\end{eqnarray}
%\end{widetext}
There we neglected those terms in second order $\sim \lambda^2$ which lead
only
to a correction of the isotropic exchange interaction.
%Now, we specify the general theory to the special Cu-O-Cu bond sketched in
%Fig.\ 1 and start with the transfer terms. In the first part of the
%calculation we neglect Hund's exchange ($\hat{H}_H$).
%We only list the relevant transfer terms for a
%coupling of $|A,0\rangle$ and $|B,0\rangle$ due to
%transfer and SO interaction.
To apply the general theory to the given situation one has to specify the
transfer terms that couple $|A,0\rangle$ and $|B,0\rangle$ due to the
different perturbations. First, we neglect Hund's exchange.
The oxygen $p_x$, $p_y$ and $p_z$ orbitals are
defined in the
local coordinate system of Cu$_A$ ($x$-$y$-$z$ in difference to the
$x'$-$y'$-$z'$ system at Cu$_B$) and the relevant transfer terms
between Cu$_A$ and the intermediate oxygen are
$t_{A0,p_x}=t_{pd}$ and
$t_{Ay,p_z}=t_{pd}/\sqrt{3}$,
where we used the ratio between $\sigma$ and $\pi$ transfer to be $1:\sqrt{3}$
which is valid with good approximation.
%The transfer $t_{Az,p_y}=t_{Ay,p_z}$
%is not important (without $\hat{H}_H$) since $|p_y \rangle$ has no
%transfer to
%$|B,0 \rangle$.
To obtain the transfer integrals of oxygen with the
Cu$_B$ orbitals we have to rotate the $|p_x'\rangle$ and $|p_y'\rangle$
orbitals to $|p_x \rangle$ and $|p_y\rangle$. Then we get the transfer
terms
\begin{eqnarray}
t_{B0,p_x}=\cos \phi \ t_{pd} \quad &,& \quad
t_{B0,p_z}=- \sin \phi \ t_{pd} \; ,  \nonumber \\
t_{By,p_x}=\sin \phi \ \frac{t_{pd}}{\sqrt{3}} \quad &,& \quad
t_{By,p_z}=\cos \phi \ \frac{t_{pd}}{\sqrt{3}} \; . 
\label{8}
\end{eqnarray}
The next step consists in calculating the exchange integrals in $\hat{H}^{ex}$.
Let us start with the isotropic exchange $J=J(0_A,0_B;0_A,0_B)$. We use
perturbation theory with respect to $\hat{H}_t$ in 4th order. There are two
possible intermediate states, the doubly occupied oxygen state $|p_x \rangle$
with an energy $2 \Delta_p + U_p$ and the doubly occupied copper state
with the energy $U_d$. Collecting all the possible paths we get
\begin{equation}
\label{9}
J=4 G b^2 \quad
\mbox{with} \quad
%b=t_{pd}^2 \cos \phi \quad
%\mbox{and} \quad
G = \frac{1}{\Delta_p^2}
\left( \frac{1}{U_d} + \frac{2}{2 \Delta_p + U_p} \right) 
\end{equation}
and $b=t_{pd}^2 \cos \phi$.
With the standard value for cuprates $t_{pd}=1.3$ eV
we obtain for BaCu$_2$Ge$_2$O$_7$ with 
$\phi=45^{\circ}$ the estimate $J\approx 95$ meV being roughly two 
times larger than the experimental value. 
The reason is most probably the insufficiency of 4th
order perturbation theory which is also known for the standard CuO$_2$
plane. For the Si-compound 
($\phi=56^{\circ}$) we estimate 60 meV, whereas
the experimental value is 24 meV.
Now, the DM vector $\vec{D}$ shall be calculated.
The two ground states $|A,0\rangle$ and
$|B,0\rangle$ are coupled by SO and $\hat{H}_t$ only via the
$|B,y\rangle$ or the $|A,y\rangle$ orbitals. Therefore, only $D^y$ is
different from zero and the corresponding exchange paths (in 4th order) are
shown in Fig.\ 3 with 
\begin{equation}
\label{10}
J(0_A,y_B;0_A,0_B)=
\frac{4 G t_{pd}^4 \cos \phi \sin \phi}{\sqrt{3}}
\end{equation}
being equal to $-J(y_A,0_B;0_A,0_B)$.
%With $L_{y0}^y=-i$ 
We find
\begin{equation}
\label{11}
D^y=8Gbc
\quad \mbox{where} \quad
c = \frac{\lambda}{\varepsilon_d} \sin \phi \frac{t_{pd}}{\sqrt{3}} \; .
\end{equation}
The corresponding canting angle
\begin{equation}
\label{12}
\Theta=\frac{|D^y|}{|J|} = \frac{2}{\sqrt{3}}
\left(\frac{\lambda}{\varepsilon_d}\right) \tan \phi
%\approx 0.05
\end{equation}
results in 0.058 for BaCu$_2$Ge$_2$O$_7$
whereas 0.033 was estimated on the basis of the
magnetization data \cite{tsukada00}.
\begin{figure}[t]
\includegraphics[width=7.5cm,angle=0]{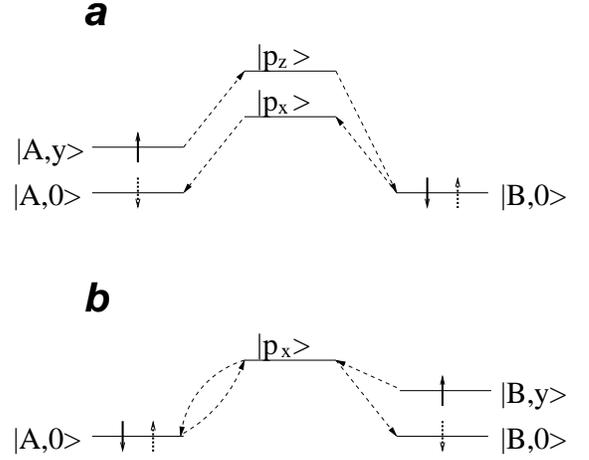}
\caption{
Exchange paths for the DM term, contributions to
$J(y_A,0_B;0_A,0_B)$ (a) and $J(0_A,y_B;0_A,0_B)$ (b). The initial state is
indicated by full arrows, the final state by dashed arrows. }
\end{figure}

The contribution in second order of the SO coupling (\ref{3}) corresponds
to a symmetric anisotropy term.
%At first we will neglect Hund's exchange, and
%we will see
%that the corresponding contribution compensates the DM term.
The two relevant
nonzero exchange terms are $J(y_A,0_B;y_A,0_B)=-J(y_A,y_B;0_A,0_B)$. The
corresponding exchange paths are very similar to those shown in Fig.\ 3. So,
the first term is characterized by an exchange between $|A,y\rangle$ and
$|B,0\rangle$ via the $|p_z\rangle$ orbital. The second, mixed exchange goes
from $|A,y\rangle$ via $|p_z\rangle$ to $|B,0\rangle$ and from $|B,y\rangle$
via $|p_x\rangle$ to $|A,0\rangle$. The result can be given as
\begin{equation}
\label{13}
J(y_A,0_B;y_A,0_B)=
4 G \left( \frac{t_{pd}^2 \sin \phi }{\sqrt{3}} \right)^2 \; .
\end{equation}
Summing both contributions we get 
\begin{equation}
\label{14}
\Gamma^{yy}=4 G c^2 \; . 
\end{equation}
Now, we see that we can combine the isotropic exchange
(\ref{9}), the DM term (\ref{11}) and the symmetric anisotropy (\ref{14})
\begin{eqnarray}
\label{15}
&&4 G \left(
(b^2-c^2) \vec{S}_A \vec{S}_B + 2 bc (\vec{S}_A \times \vec{S}_B )_y
+ 2 c^2 S_A^yS_B^y \right)
\nonumber \\
&&=\tilde{J} \tilde{\vec{S}}_A \tilde{\vec{S}}_B
\end{eqnarray}
to an isotropic exchange of canted spins $\tilde{\vec{S}}_{A/B}$ with the
canting angle $\Theta$ (\ref{12}) \cite{shekhtman92,aharony98}.
In the zigzag
chain one can subsequently
rotate all the spins. As a consequence, we would obtain a Hamiltonian which is
rotationally invariant in spin space and
no crystallographic direction would
be preferred.
That is exactly the compensation
as proposed in Refs.\ \cite{kaplan83,shekhtman92}.
%
%a la Aharony \cite{aharony}.
%Please note that the second contribution to $\Gamma^{yy}$ (\ref{6})
%proportional to the exchange interaction $J(m_A,m'_B;0_A,0_B)$ (where the SO
%coupling acts on both ions) contributes equally well to the compensation like
%the first term. But the second term was considered to be of less importance
%and was not explicitly given in the text book of Yosida \cite{yosida}.
Now, we are going to show that Hund's exchange interaction (\ref{4}) leads to
a ``residual anisotropy''.
These symmetric anisotropy terms exist already for $\phi=0$.
The anisotropy terms (\ref{11}) and (\ref{14}) can be eliminated by the
rotation (\ref{15}) and only the additional contributions due to
(\ref{4}) are considered further on. Starting with Hund's exchange at oxygen
$J_{Hp}$, new exchange integrals become 
possible in 4th order of $\hat{H}_t$ and
first order of $\hat{H}_H$. For example, the exchange paths for
$J^p(z_A,0_B;z_A,0_B)$ are realized due to the transfer term $t_{Az,p_y}
(=t_{Ay,p_z})$. It turns out that it is identical to
$J^p(y_A,0_B;y_A,0_B)=J^{(p)}$:
\begin{equation}
\label{16}
J^{(p)}=
- \frac{8}{3} \frac{t_{pd}^4 \cos^2 \phi }{\Delta_p^2}
\frac{J_{Hp}}{(2\Delta_p+U_p)^2} \; . 
\end{equation}
We have no contribution via $|A,x\rangle$ or $|B,x\rangle$, and therefore,
$\Gamma^{xx}=0$.
Next, we consider Hund's exchange at copper $J_{Hd}$. Then the transfer
terms $t_{By,p_z}=t_{B1,p_z}=\cos \phi t_{pd}/\sqrt{3}$ become important.
%One
%contribution is
%shown in Fig.\ 3b. An analogous one goes via $|B,y\rangle$. Together one
One obtains
\begin{equation}
\label{17}
J^{(d)}=
- \frac{4}{9} \frac{t_{pd}^4 \cos^2 \phi }{\Delta_p^2}
\frac{J_{Hd}}{U_d^2} \; . 
\end{equation}
Both terms (\ref{16},\ref{17}) together result in the
``residual anisotropy'' contribution
\begin{equation}
\label{18}
\Gamma^{yy}=\frac{\Gamma^{zz}}{4}=
\frac{\lambda^2}{\varepsilon_d^2} \left(
J^{(d)} + J^{(p)} \right)
\end{equation}
of one plaquette.
To connect the cluster of Fig.\ 2 with the real crystallographic structure
(see Fig.\ 1 in \cite{tsukada99}), one has to rotate the coordinate axes
$\vec{e}_z=(z_a,z_b,z_c)=(0.84,0.38,0.40)$ and
$\vec{e}_y=(y_a,-y_b,-y_c)=(0.48,-0.85,-0.21)$ (and also $\vec{e}_{z'}$
and $\vec{e}_{y'}$) to the crystallographic ones. We obtain the
diagonal elements
\begin{equation}
\label{19}
\Gamma^{\nu\nu}=\frac{\Gamma^{yy}}{2}
\left( 4 z_{\nu}^2 + y_{\nu}^2 \right)
\end{equation}
with $\nu=\{a,b,c\}$ and only one non-zero off-diagonal element $\Gamma^{ab}$.
With the characteristic parameters $J_{Hp}=0.4$ eV and $J_{Hd}=1$ eV and
setting $\Gamma^{cc}$ to zero, we get 
$\Gamma^{aa}=-2 \ \mu$eV and $\Gamma^{bb}=-0.5 \ \mu$eV.
These numbers are extremely small, but we have checked that they are still
larger than the classical dipole-dipole interaction.
For an antiferromagnetic arrangements of spins along the chain, the preferred
spin direction is the $c$-direction which is indeed the case for
BaCu$_2$Ge$_2$O$_7$ \cite{tsukada00} and BaCu$_2$Si$_2$O$_7$
\cite{zheludev01}.
%\begin{figure}[h]
%\includegraphics[width=8.5cm,angle=0]{f3.eps}
%\caption{
%Exchange paths for the ``residual anisotropy'' contribution due to
%$J_{pH}$ (a) and due to $J_{dH}$ (b). }
%\end{figure}
From the theoretical analysis follows that we can expect for
BaCu$_2$Si$_2$O$_7$ an easy $c$-axis, a middle $b$-axis and a hard
$a$-axis. The diagonal anisotropy energies (\ref{19}) are directly
related to the two gaps seen in Fig.\ 1 (and also in neutron
scattering \cite{kenzelmann01}) according to the standard formulas
$\nu_{LF}$=$\sqrt{2H_{A1}H_E}$ and
$\nu_{HF}$=$\sqrt{2H_{A2}H_E}$ with $H_E=z J S$ ($S$ - spin, $z$ -
number of nearest neighbors) and $H_{A1}=2 z |\Gamma^{b b}| S$,
$H_{A2}=2 z |\Gamma^{a a}| S$ \cite{remark4}. This yields 0.44 meV
and 0.22 meV in reasonable agreement with the experimental values
0.36 meV and 0.21 meV. 

In conclusion, our theory is applicable to folded CuO$_3$ chains and gives 
the isotropic exchange, the DM term and the spin canting angle, the 
compensating symmetric exchange and the ``residual anisotropy'' due to 
Hund's coupling in good accord with the available experimental data 
of BaCu$_2$Ge$_2$O$_7$ and BaCu$_2$Si$_2$O$_7$ and especially with 
the two gaps measured by AFMR for the latter compound. Possible 
differences between the two compounds might be caused by 
different interchain couplings which were not included in the present study.

%Concluding we remark that the theoretical estimate spans an extreme 
%range of energies (from 100 meV to 1 $\mu$eV) but is in good accord 
%with the experimental data and especially with the two gaps measured by 
%AFMR. 
%The heavy direction can be understood by a simple rule:
%normally, in cuprates with an antiferromagnetic nearest neighbor 
%coupling, the heavy direction is perpendicular to the plaquettes. 
%And in the given compound 
%the plaquette normals have a
%maximal projection for the crystallographic $a$-axis which explains the
%observed behavior. 
%The most important question which was not touched in
%the present study concerns the possible relevance 
%of interchain couplings.

We thank J.-L.\ Richard for discussions and  
NATO (PST.CLG.976416) 
and DFG (UKR 113/49/0-1) for support.  


\begin{thebibliography}{30}
\bibitem{remark5}
A more detailed discussion of the AFMR in BaCu$_2$Si$_2$O$_7$
will be published elsewhere.
\bibitem{remark4}
By inspection of the equations of motion it is easily seen that the
anisotropy in $a$-direction influences the magnetic susceptibility in
$b$-direction and vice versa.
\end{thebibliography}
\end{document}